\documentclass[a4paper,fleqn,usenatbib]{mnras}

\usepackage[T1]{fontenc}
\usepackage{ae,aecompl}

\usepackage{graphicx}	
\usepackage{amsmath}	
\usepackage{amssymb}	

\title[Intra-group medium of a $z=0.28$ galaxy group]{Probing the intra-group medium of a $z=0.28$ galaxy group}

\author[R. Bielby]
{R. Bielby$^{1}$\thanks{E-mail: richard.bielby@durham.ac.uk (RMB)}, N. H. M. Crighton$^2$, M. Fumagalli$^{1,3}$, S. L. Morris$^{1}$, J. P. Stott$^{4,5}$,
	\newauthor N. Tejos$^{6,7}$, S. Cantalupo$^{8}$
\\
$^1$ Centre for Extragalactic Astronomy, Durham University, South Road, Durham DH1 3LE, UK\\
$^2$ Centre for Astrophysics and Supercomputing, Swinburne University of Technology, Hawthorn, Victoria 3122, Australia\\
$^3$ Institute for Computational Cosmology, Durham University, South Road, Durham DH1 3LE, UK\\
$^4$ Department of Physics, University of Oxford, Denys Wilkinson Building, Keble Road, Oxford OX1 3RH, UK\\
$^5$ Department of Physics, Lancaster University, Lancaster LA1 4YB, UK\\
$^6$ Millennium Institute of Astrophysics, Casilla 36-D, Santiago, Chile\\
$^7$ Instituto de Astrof\'isica, Pontificia Universidad Cat\'olica de Chile, Vicu\~na Mackenna 4860, Santiago, Chile\\
$^8$ ETH  Zurich,  Institute  of  Astronomy,  Wolfgang-Pauli-Str. 27, 8093 Zurich, Switzerland
}

\date{Accepted XXX. Received YYY; in original form ZZZ}

\pubyear{2016}

\begin{document}
\label{firstpage}
\pagerange{\pageref{firstpage}--\pageref{lastpage}}
\maketitle

\begin{abstract}
We present new MUSE observations of a galaxy group probed by a background quasar. The quasar sightline passes between multiple $z=0.28$ galaxies, whilst showing at the same redshift low ionised metal line species, including Ca~{\sc ii}, Mg~{\sc i}, Mg~{\sc ii} and Fe~{\sc ii}. Based on the galaxy redshifts measured from the MUSE data, we estimate the galaxies to be part of a small galaxy group with a halo mass of $\approx6\times10^{12}$~M$_{\odot}$. We use the MUSE data to reveal the two dimensional dynamical properties of the gas and stars in the group galaxies, and relate these to the absorber kinematics. With these data we consider a number of scenarios for the nature of the gas probed by the sightline absorbers: a co-rotating gas halo associated with a single galaxy within the group; outflowing material from a single group member powered by recent star-formation; and cool dense gas associated with an intra-group medium. We find that the dynamics, galaxy impact parameters, star-formation rates, and the absorber strength suggest the cool gas can not be clearly associated with any single galaxy within the group. Instead we find that the observations are consistent with a superposition of cool gas clouds originating with the observed galaxies as they fall into the group potential, and are now likely in the process of forming the intra-group medium.
\end{abstract}




\section{Introduction}

The interplay between galaxies and the gaseous media that surround them is a crucial element in the study of galaxy formation, providing insights into both the infall of material onto galaxies and the extent and impact of galactic outflows. Observing the baryonic material surrounding galaxies is primarily performed (at least beyond modest redshifts) using absorption line studies, either in the line of sight to background QSOs (quasi-stellar object) or galaxies. Much work has been performed on both the small scales, linking absorption systems to individual galaxies \citep[e.g.][]{2011ApJ...733..105K,2012ApJ...760L...7K,2012MNRAS.426..801B,2013Sci...341...50B,2013ApJ...777...59T,2013ApJ...763..148S,2013MNRAS.436.2650P,2013ApJ...776..115N,2015ApJ...812...83N,2016ApJ...818..171N,2014ApJ...792....8W,2015MNRAS.446.3178F}; and on statistical scales: analysing the large scale distribution of gas around galaxies \citep[e.g.][]{adelberger03,adelberger05,2006MNRAS.367.1251R,2006MNRAS.367.1261M,2009ApJ...701.1219C,2010MNRAS.402.2520S,2011MNRAS.414...28C,2012ApJ...750...67R,2012ApJ...751...94R,2014MNRAS.437.2017T,2014MNRAS.442.2094T,2014MNRAS.445..794T,2015MNRAS.450.2067T,2016MNRAS.460..590F,2016arXiv161009144B}.

An important, but poorly constrained, element in connecting absorbers with individual galaxies however, is the galaxies' place in the large scale structure of the Universe. Many studies focus on so-called field galaxies with little or no quantification of the nearby galaxy population. Moreover, much of the star forming galaxy population resides in groups and this element is rarely considered in the analysis of absorption line systems, especially at high redshift. Absorbers are often treated as part of a distinct galaxy halo, whereas in reality the distribution of gas around galaxies is more complex and it may not be clear or reasonable to constrain where the halo of one galaxy finishes and the next begins \citep[e.g.][]{1994Natur.372..530Y,2002ASPC..254...72M}.

In terms of absorption of background light by gas within galaxy groups, \citet{1994ApJ...427..696M} suggested that a significant fraction of H~{\sc i} absorption systems (with column densities above of $\sim10^{13}$~cm$^{-2}$) are the result of gaseous pressure-confined tidal debris, finding comparable space densities between low-column density H~{\sc i} absorbers and galaxy groups in the local Universe. Such absorption studies are complemented by H~{\sc i} emission maps of nearby galaxies and small groups \citep[e.g.][]{1999AJ....117..811H}, which show significant clumps of H~{\sc i} gas apparently tidally or ram-pressure stripped from the surrounding galaxies.

Further work has been performed analysing the place of H~{\sc i} gas (over a range of column densities) within the large-scale-structure, with \citet{2002ApJ...565..720P}, for example, looking at the relationship between Ly{$\alpha$} absorbers and voids and super-clusters. Through a clustering analysis, they reported that Ly$\alpha$ absorbers (with column densities of $10^{13.2}$ to $10^{15.4}$~cm$^{-2}$) cluster with galaxies, but more weakly than galaxies cluster with each other. As such, \citet{2002ApJ...565..720P} conclude that such Ly$\alpha$ absorption systems are primarily associated with the large-scale structures of galaxies, i.e. filaments and groups, rather than individual galaxies \citep[see also][]{2014MNRAS.437.2017T}.

Continuing such considerations of the relationship between absorbers and the large scale structure, \citet{2010MNRAS.402.1273C} found evidence that galaxy groups trace Mpc-scale structures that appear in both H~{\sc i} gas and galaxies based on the study of a triplet of closely separated QSO sightlines. Additionally, \citet{2012MNRAS.425..245T} found peaks in the Ly{$\alpha$} absorbers distribution around the edges of large voids: i.e. tracing the filamentary structures enclosing such voids (a topic further developed in \citealt{2016MNRAS.455.2662T}).

Consistent with the column densities of H~{\sc i} emission maps discussed above, \citet{2000ApJ...543L...9C} presented the detection of a Lyman limit system (LLS) apparently associated with a pair of galaxies at $z=0.167$. Significantly, the reported LLS corresponded in velocity space to an O~{\sc vi} absorption system, which the authors claimed as tracing warm gas within the intra-group medium of this small galaxy group. Given that this detection of O~{\sc vi} absorption fits with the gas temperatures expected for the intra-group medium of typical galaxy groups (i.e. $T\sim10^5$~K), much interest has since been placed in tracing the potentially huge reservoirs of gas within galaxy groups using such absorption systems. Indeed, \citet{2006ApJ...643..680P} find O~{\sc vi} absorption arises in a diverse set of galactic environments including the halos of individual galaxies, galaxy groups, filamentary-like structures, and also regions devoid of luminous galaxies, albeit based on a relatively small sample of galaxies and absorbers. Similarly, \citet{2009ApJ...701.1219C} find that tidal debris in groups/galaxy pairs may be principally responsible for observed O~{\sc vi} absorbers (although this is somewhat in contrast to other findings, e.g. \citealt{2011Sci...334..948T}).

\citet{2014ApJ...791..128S} presented a study of absorber measurements in quasar sightlines, observed with HST/COS, passing through 14 galaxy groups. Their far-UV spectra produced 14 warm ($T\geq10^5$~K) absorption systems consisting of broad Ly$\alpha$ and broad O~{\sc vi} absorption, which were each found to be coincident with spiral-rich groups or cluster outskirts. Their analysis tentatively favours the absorbers to be associated with the intra-group/cluster medium rather than the halos of individual galaxies, albeit with some systematic uncertainties. Supporting this finding, \citet{2015MNRAS.449.3263J} analyse the effect of environment on their analyses of O~{\sc vi}, finding that galaxies with nearby neighbours exhibit a modest increase in O~{\sc vi} covering fraction compared to isolated galaxies suggest that environmental effects play a role in distributing heavy elements beyond the enriched gaseous haloes of individual galaxies, whilst \citet{2016MNRAS.458..733P} identify O~{\sc vi} and broad Ly~$\alpha$ absorbers associated with $z=0.4$ galaxy group.

However, although there has been much focus on O~{\sc vi} as a tracer of intra-group gas, in reality the intra-group medium is likely a complex mix of phases incorporating cool gas, potentially in clumpy structures or clouds, as well as the warmer gas phases \citep{2016ApJ...830...87S}. Mg~{\sc ii} as a doublet observed at optical wavelengths represents an excellent tracer of cool gas and as such it is an important and complementary diagnostic to add to the O~{\sc vi} studies discussed above. Only a small number of Mg~{\sc ii}-galaxy group associations have been reported in detail however: \citet{2006MNRAS.368..341W} detected Mg~{\sc ii} absorption coincident with a galaxy group at $z=0.666$; \citet{2010MNRAS.406..445K} reported a strong Mg~{\sc ii} absorber ($W_r(2796)=1.8$~\AA) coincident with a galaxy group with 5 observed members at $z=0.313$; and \citet{2013MNRAS.432.1444G} investigated a large equivalent width (EW) Mg~{\sc ii} absorber detected in close proximity to a galaxy group at $z\sim0.5$. The latter study concluded that the absorption system is most likely associated with the intra-group medium, with the authors favouring a scenario whereby the absorption originates in tidal debris. To give a sense of the number of potential Mg~{\sc ii} absorber-galaxy group associations, we note that from a sample of $\sim200$ Mg~{\sc ii} absorbers and nearby galaxies, \citet{2013ApJ...776..114N} found that $\approx40$ appeared to be associated with group galaxies or were ambiguous in what galaxy they may be associated with. However, due to the complexity of disentangling the relationship between absorbers found in a group environment and the group galaxies, most studies have taken the approach of either removing the group absorbers from their analysis \citep[e.g.][]{2013ApJ...779...87C,2013ApJ...776..115N,2015ApJ...812...83N} or simply taking the group galaxy with the smallest impact parameter \citep[e.g.][]{2016ApJ...833...39S} as the absorber `host'.

Here, we present serendipitous MUSE observations of a galaxy group observed at $z=0.2825$ in the foreground of a bright $z=1.71$ quasar and coincident with significant metal line absorption features. MUSE represents an excellent probe of galaxy-group structures around QSO sightlines, providing as it does a blind survey over an area of $1'\times1'$ \citep[e.g.][]{2016MNRAS.462.1978F,2017MNRAS.464.2053P}. Our observations are nominally part of the QSO Sightline and Galaxy Evolution (QSAGE) survey, which aims to provide deep galaxy data around HST/STIS observed QSO sightlines, primarily using the WFC3 G141 grism (HST Program 14594, PI: R. Bielby). In Section~\ref{sec:observations} of this paper, we present our MUSE observations and give an overview of the available ancillary galaxy data and the sightline spectral data. In Section~\ref{sec:analysis}, we present an analysis of the galaxy and group properties alongside an analysis of the sightline absorbers. In Section~\ref{sec:discussion}, we discuss the relationship between the group galaxies and the absorbers and the implications for the nature of the group environment. We present our conclusions in Section~\ref{sec:conclusions}. We assume a Planck 2013 cosmology \citep{2014A&A...571A..16P} throughout. All quoted distances are in proper coordinates unless stated otherwise.

\section{Observations}
\label{sec:observations}

\subsection{MUSE observations}

\subsubsection{Overview}

The MUSE instrument provides integral field spectroscopy of a contiguous square field of view with a width of $1'$. MUSE provides a spatial pixel scale of $0.2''$~pixel$^{-1}$, whilst the spectral resolution gives a resolution element of $\Delta\lambda\approx3$~\AA. This is over a wavelength range of 4800~\AA~$<\lambda<9300$~\AA~(in the nominal mode used here).

The MUSE data presented here were taken as part of ESO project 094.B-0304(A) aimed at identifying galaxy counterparts to a sub-DLA system in the HE0515-4414 sightline at $z=1.15$. Out of 5 intended 1 hr observing blocks, only 2 were attempted during the allotted semester, with the first being out of the condition constraints (due to poor seeing) and the second being aborted due to high wind. Here we present results from the successfully taken observing block, albeit at poorer seeing than required for the original science goals.

The data was observed 6th Feb 2015 with 2 integrations, the first of duration 1458s and the second 548s (cut short due to worsening conditions). Given the increasingly poor atmospheric conditions for the second exposure, we solely use the first, uninterrupted, exposure for all the analysis that follows.

\subsubsection{MUSE data reduction}

The muse data reduction was performed using a combination of recipes from the ESO MUSE pipeline \citep[v1.2.1;][]{2014ASPC..485..451W}, the {\sc CubExtractor} package ({\sc CubEx}, version 1.5; Cantalupo, in prep.), and in-house Python codes.

We created the master bias, master flat and a wavelength solution using the MUSE pipeline. The illumination correction was then calculated using the adjacently taken flat-field. These calibrations were applied to the science exposures and the standard stars used for spectro-photometric flux calibration. The individual exposures were transformed into data-cubes with a common reference grid (voxels of 1.25~\AA\ in the spectral direction and $0.2''$ in the spatial direction) and barycentric corrections were applied.

A two-stage optimisation of the data was then performed using the {\sc CubEx} package. We first applied a flat-field correction and sky-subtraction using the {\sc CubeFix} and {\sc CubeSharp} routines within {\sc CubEx} \citep[e.g.][]{2016ApJ...831...39B} and from the resultant cubes created collapsed white-light images. Using these images, we identified sources and created source masks and then re-ran the {\sc CubEx} flat-field correction and sky-subtraction with these source masks applied.

For the analysis presented in this paper, the wavelength axis was transformed from air to vacuum (matching the QSO sightline spectra described later). Our final white-light image (i.e. produced from collapsing the cube along the wavelength direction) is shown in Fig.~\ref{fig:musefov}. The narrow horizontal bands visible in the image are the result of the boundaries between the four banks of individual detector elements in the MUSE instrument. The data-cube has seeing of $1.2''$ and 3$\sigma$ depth of $f_{3\sigma}=16\times10^{-18}$~erg~cm$^{-2}$~s$^{-1}$~\AA$^{-1}$ (based on a 3D aperture with FWHM of $1.2''$ and $3.0$~\AA\ in the spatial and spectral directions respectively).

\subsubsection{Galaxy identification}

In order to identify sources within the final cube, we ran {\sc SExtractor} \citep{sextractor} on the white-light image. We removed clearly spurious objects via visual inspection (e.g. caused by noise spikes, chip defects, imperfections in the background subtraction around chip edges) from the resulting catalogue and proceeded to extract spectra and identify individual objects within the field. The primary features used to visually identify object redshifts, given the wavelength range and redshifts covered, were [O~{\sc ii}]~3727\AA, Ca~K 3933\AA\ and H~3869\AA\ absorption, H~$\beta$ 4861\AA, [O~{\sc iii}]~4959\AA, [O~{\sc iii}]~5007\AA, H~$\alpha$~6562\AA, and [N~{\sc ii}]~6583\AA. Redshifts were obtained for galaxies showing nebular emission by fitting the [O~{\sc ii}], H~$\beta$, [O~{\sc iii}], and H~$\alpha$ lines individually with a Gaussian (or double Gaussian in the case of [O~{\sc ii}]) and taking the average. For passive galaxies with little or no nebular emission, we simultaneously fit the Ca H and K absorption lines, again with Gaussian profiles. With this initial solution, we then fit the absorption line galaxy spectra using the Penalized Pixel-Fitting code described by \citet{2004PASP..116..138C} and \citet{2017MNRAS.466..798C} in conjunction with the spectral templates of \citet{2006MNRAS.371..703S}. Redshift uncertainties are derived using Monte-Carlo realisations of the data incorporating the measured noise on the galaxy spectra and are given alongside the estimated redshifts in Table~\ref{tab:catalogue}.

\begin{figure*}
	\includegraphics[width=12cm]{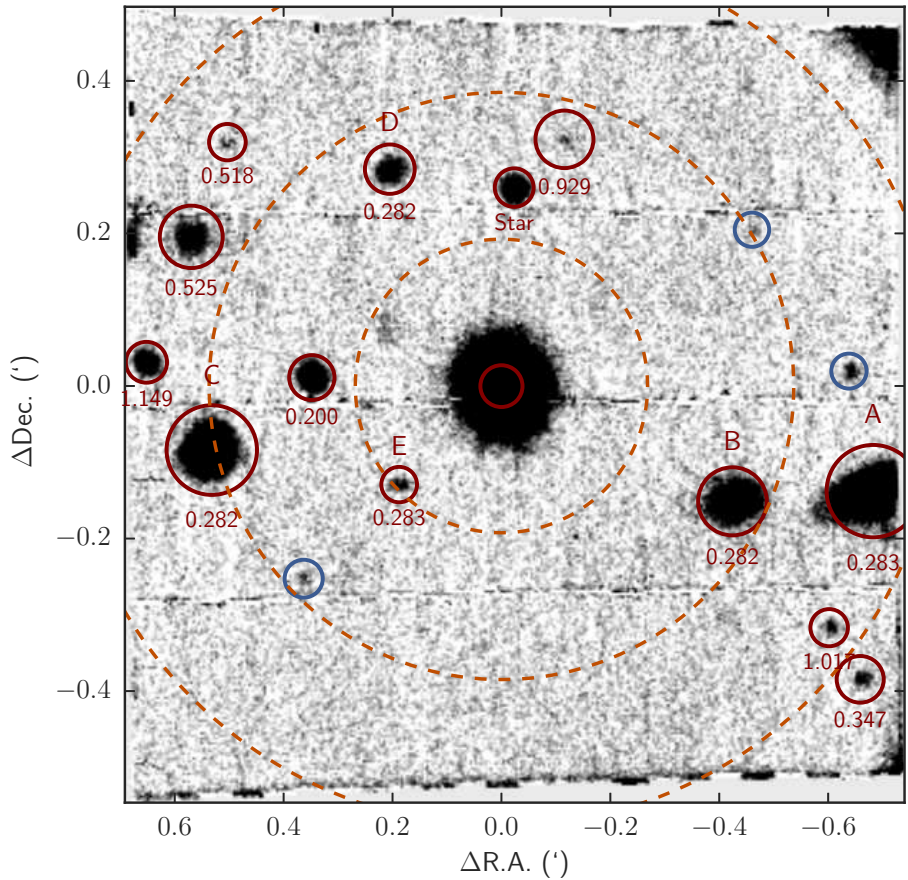}
    \caption{White-image of the MUSE cube centred on the bright $z=1.7$ quasar HE0515-4414. The axes show the offset from the central quasar in arc-minutes. Detected sources in the MUSE data are circles, whilst those with identified redshifts are labelled. The background quasar lies between five identified foreground galaxies forming a galaxy group at $z=0.2825$. The dashed concentric rings centred on the QSO show impact parameters of $b=50$~kpc, $b=100$~kpc, and $b=150$~kpc (at $z=0.28$).}
    \label{fig:musefov}
\end{figure*}

The resultant galaxy identifications are shown in Fig.~\ref{fig:musefov}, where each identified source is circled and (where one was successfully obtained) its redshift given. The sources are also listed in Table~\ref{tab:catalogue}. 

\begin{table*}
\caption{Catalogue of objects identified in the MUSE data centred on the bright QSO HE0515-4414.}
\label{tab:catalogue}
\begin{tabular}{lcccccccc}
\hline
ID & R.A.           & Dec.                & $R_{e}$ & Redshift & Class & $R$ & $M_{\rm B}$ & $M_{\rm R}$\\
   & \multicolumn{2}{c}{(J2000)} & ($''$)  &          &       &(mag)&    (mag)   & (mag) \\
\hline
 J051709-441111                   &  79.287842 & -44.186323 &  0.56 & ---    & ---      &  23.41& ---   & ---    \\ 
 J051706-441043                   &  79.274123 & -44.178698 &  0.50 & ---    & ---      &  24.12& ---   & ---    \\ 
 J051705-441054                   &  79.271154 & -44.181787 &  0.52 & ---    & ---      &  22.45& ---   & ---    \\ 
 J051708-441040                   &  79.281399 & -44.177778 &  0.57 & 0.0000 & M-star   &  20.18& ---   & ---    \\ 
 J051709-441055                   &  79.287592 & -44.181929 &  0.66 & 0.1997 & Em       &  20.12& -19.70& -20.00 \\ 
 J051708-441039                   &  79.285206 & -44.177375 &  0.73 & 0.2821 & Em       &  20.82& -19.81& -20.27 \\ 
 MRSS252-053745                   &  79.290654 & -44.183513 &  1.32 & 0.2823 & Em       &  18.73& -22.02& -22.26 \\ 
 MRSS252-054266                   &  79.274720 & -44.184633 &  1.00 & 0.2825 & Ab       &  19.07& -21.44& -22.03 \\ 
 MRSS252-054422                   &  79.270413 & -44.184410 &  1.36 & 0.2826 & Ab       &  18.72& -21.83& -22.47 \\ 
 J051708-441103                   &  79.284920 & -44.184267 &  0.52 & 0.2835 & Em       &  22.74& -18.13& -17.94 \\ 
 J051705-441119                   &  79.270810 & -44.188522 &  0.69 & 0.3474 & Em       &  22.58& -18.74& -19.06 \\ 
 J051710-441036                   &  79.290177 & -44.176785 &  0.54 & 0.5185 & Em       &  23.47& -19.20& ---    \\ 
 J051710-441044                   &  79.291282 & -44.178854 &  0.92 & 0.5253 & Ab       &  20.36& -22.34& ---    \\ 
 J051707-441036                   &  79.279861 & -44.176723 &  0.85 & 0.9289 & Em       &  23.84& -20.59& ---    \\ 
 J051705-441115                   &  79.271760 & -44.187394 &  0.55 & 1.0174 & Em       &  23.01& -21.88& ---    \\ 
 J051710-441054                   &  79.292655 & -44.181594 &  0.60 & 1.1492 & Em       &  20.67& -24.39& ---    \\ 
 HE0515-4414                      &  79.281790 & -44.182116 &  0.62 & 1.7176 & QSO      &  14.65& ---   & ---    \\ 

\hline
\end{tabular}
\end{table*}


\subsection{QSO sightline}

In this work we use archival UVES spectra of the the HE05015-4414 QSO, taken from the ESO archive (ESO programs 66.A-0212, 076.B-0055 and 082.A-0078). These data were reduced, calibrated, and combined as described by \citet{2017MNRAS.464.3679K}. The data cover the wavelength range $3,050<\lambda<3,910$~\AA, with a total integration time of $\approx78,000$~s and a resolution of $R\approx60,000$. After correcting the reduced spectra for helio-centric velocity offsets and converting wavelengths from air to vacuum, we combine the individual exposures using a weighted mean taking into account the pipeline error estimates. The combined spectrum has a median signal-to-noise of $S/N=63$.

We also note that the quasar was observed using CASPEC by \citet{2000A&A...363...69D} in  October  1996 and November 1997. These observations cover the spectral range 3,530 to 8,570~\AA\ with $R\sim21,000-29,000$, and have a typical signal-to-noise ratio of $\approx40$. \citet{2000A&A...363...69D} performed a Voigt profile analysis of absorbers along the line of sight, which we refer to in our analysis. We update the CASPEC analysis using the above UVES data (see section~\ref{sec:absfeat}), which has both higher resolution and signal-to-noise.

\subsection{Ancillary data}

\subsubsection{FORS1 $V$-band imaging}

As part of a polarimetry study of absorption lines in the QSO spectrum (ESO program 072.B-0699), 15~s FORS1 $V$-band imaging was taken as part of the target acquisition. We reduced this data, performing bias subtraction and flat fielding with the nightly calibrations. The acquisition frames consist of $3\times5$~s exposures taken on 16th and 28th December 2003 and cover an area of $6.9'\times6.8'$ centred on the QSO.

\begin{figure*}
	\includegraphics[width=16cm]{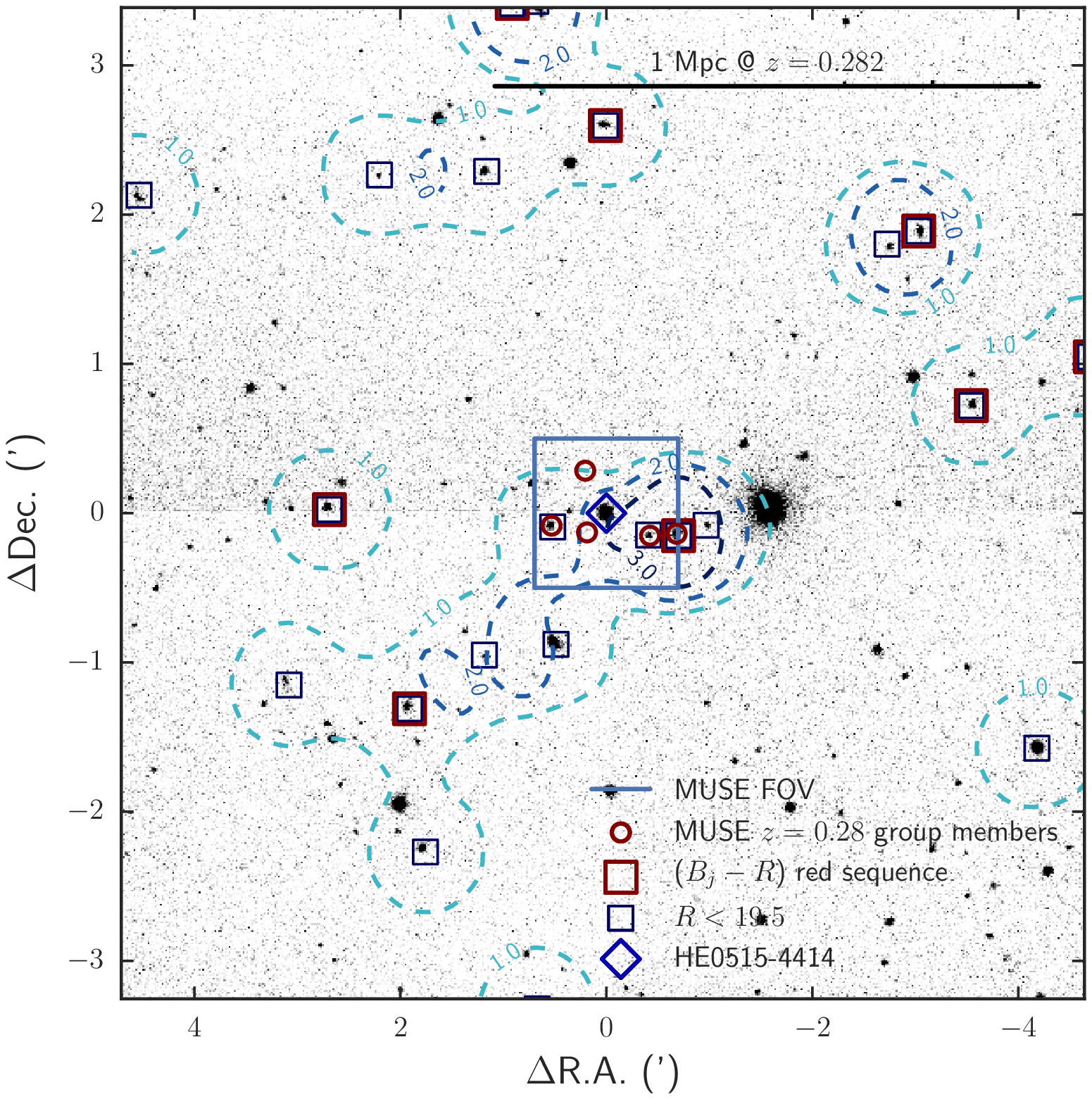}
    \caption{FORS1 $V$ band image of the region around the QSO HE0515-4414. The field of view of the MUSE data presented in this paper is shown by the $1'$ width square at the centre of the image. Sources identified within the MUSE data at $z\approx0.28$ are marked with circles, whilst the quasar is marked with a diamond. Objects identified with $(B_{\rm j}-R)$ colours consistent with being passive at $z\approx0.28$ are marked with large red squares and the smaller blue squares denote galaxies detected in the MRSS with $R<19.5$. The brightest $z=0.2825$ group member, MRSS~252-054422 (galaxy A), is marked by both a circle and the squares, falling as it does in all three categories. The dashed contours show the derived density map, with the highest density region in the field of view found to be centred on MRSS~252-054422.}
    \label{fig:ancillary}
\end{figure*}

Our stack of the three sub-exposures shows a seeing level of $1.3''$ with a $3\sigma$ magnitude limit of $V=21.5$ (based on a $0.6''$ diameter detection aperture). This stacked image is shown in Fig.~\ref{fig:ancillary}.

\subsubsection{Sky survey data}

In order to evaluate the wider area around our observations, we use data from the Automatic Plate Measuring machine Galaxy Survey \citep[APM,][]{1990MNRAS.243..692M} and the Muenster Red Sky Survey \citep[MRSS,][]{2003JAD.....9....1U}. The APM survey is complete to $B_{\rm j}\leq20.5$ (90\% completeness limit).

The MRSS data are complete to $R=19.0$~mag for extended sources at the 80\% level \citep[c.f. Fig. 4.4 of][]{2003JAD.....9....1U}. Star/galaxy separation has been performed on the MRSS sources by \citet{2003JAD.....9....1U} based on a combination of effective radius, apparent magnitude and central intensity, and in this work we use only sources classed as galaxies in MRSS. Galaxies detected by MRSS with $R<19.5$~mag (i.e. to $\approx50\%$ completeness) are highlighted by squares in Fig.~\ref{fig:ancillary}.

\section{A galaxy group at a redshift of 0.2825}
\label{sec:analysis}

\subsection{Galaxy group}

\begin{figure*}
	\includegraphics[width=\textwidth]{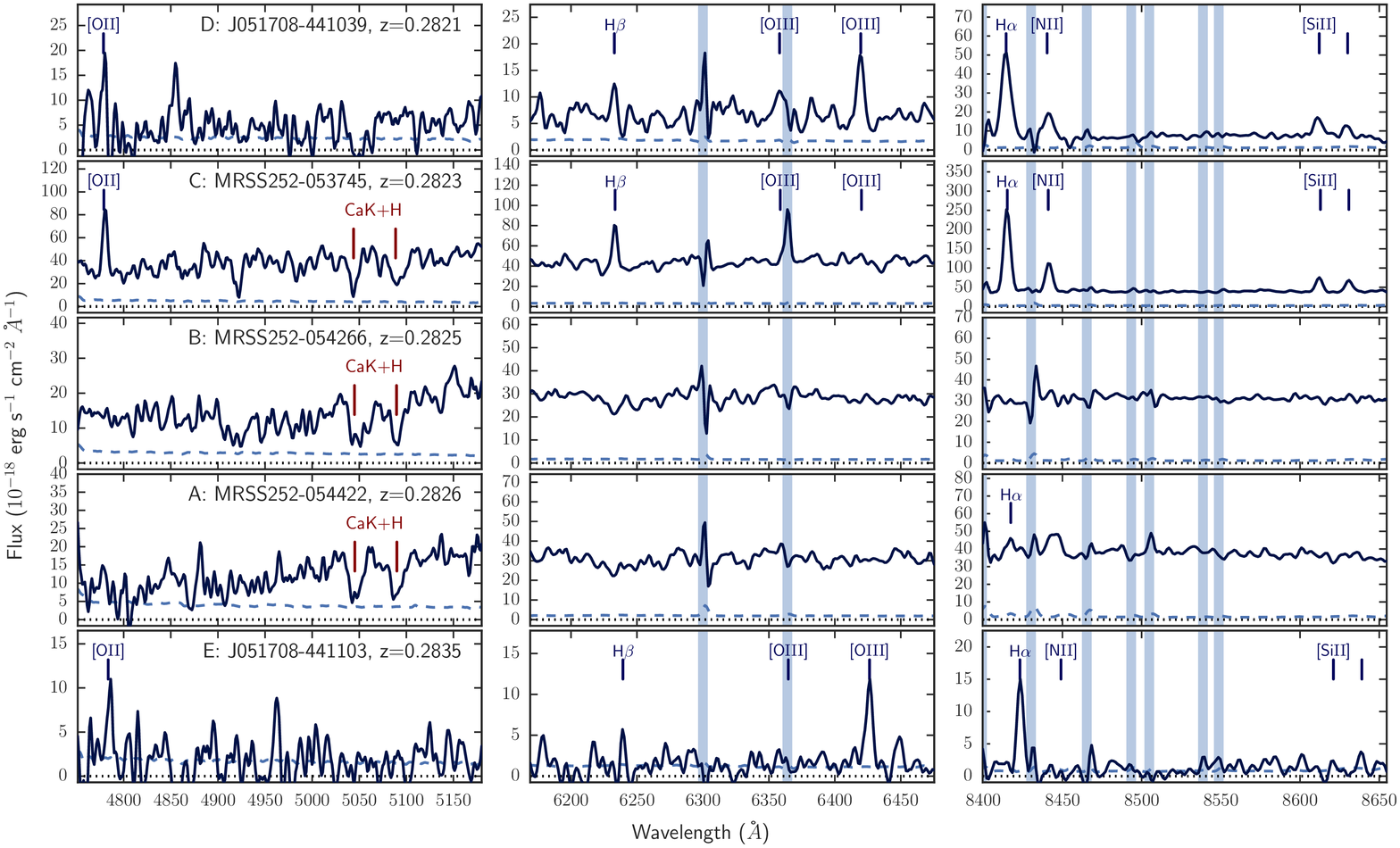}
    \caption{Extracted MUSE observed frame spectra for the five $z=0.28$ group galaxies. The left panel in each row is centred on the wavelength range of [O~{\sc ii}] and Ca H K, the centre panel on the wavelength range of H~$\beta$ and [O~{\sc iii}] and the right hand panel covers H~$\alpha$, [N~{\sc ii}] and Si~{\sc ii} emission. The solid curve in each panel shows the measured flux, whilst the dashed line shows the estimated uncertainty on that flux. The positions of key spectral features are marked. The vertical shaded regions in the central column of plots show the position of prominent atmospheric emission lines \citep{1996PASP..108..277O}.}
    \label{fig:z0.28spec}
\end{figure*}

From our MUSE observations, we find five galaxies within a tight redshift range of $z=0.2825\pm0.0006$ and separated by $\lesssim0.9'$ ($\lesssim300$~kpc). The MUSE spectra for these five galaxies, covering prominent spectral features, are shown in Fig.~\ref{fig:z0.28spec}. We denote the five group members as galaxies A, B, C, D and E as indicated in Fig.~\ref{fig:musefov} and Table~\ref{tab:galprops}.

Given the $z=0.28$ galaxies observed with MUSE are spread across the full range of R.A. covered by the MUSE data, it is reasonable to expect that further associated galaxies are present outside of the field of view. Indeed, from the MUSE data alone, it is not absolutely clear whether this is a galaxy group or the outskirts of a cluster. We therefore turn to the ancillary data around this field to clarify the nature of the observed $z=0.28$ structure. If the structure is a cluster, we would of course expect to see a large number of galaxies within scales of $\lesssim1~$Mpc with comparable colours to the passive members identified in the MUSE observations.

Firstly, a visual inspection of the FORS $V$ band imaging, shown in Fig.~\ref{fig:ancillary}, shows no obvious evidence of a strong cluster-like over-density (i.e. many galaxies across scales of $\gtrsim1$~Mpc or $\gtrsim3.8'$ at $z=0.28$) around the confirmed $z=0.28$ galaxies (marked with circles). Following \citet{2011MNRAS.416.2041M} and \citet{2016MNRAS.455..618F}, we quantify the galaxy density field by mapping the density distribution from the MRSS catalogue with a magnitude cut of $R<19.5$ (corresponding to $\approx50\%$ completeness limit of the MRSS data and equivalent to $\approx1.8~L^\star$ at $z=0.28$). Galaxies brighter than this limit are indicated by squares in Fig.~\ref{fig:ancillary}. We use a cell size of radius $0.5'$, equivalent to $\approx150$~kpc at $z=0.28$. The resultant density map is shown by the contours in Fig.~\ref{fig:ancillary}, with the labels giving the galaxy density in units of arcmin$^{-2}$. The average galaxy density at $R<19.5$ across the area is $0.26\pm0.07$~arcmin$^{-2}$, whilst the peak density within the galaxy group region is $3.0\pm1.7$~arcmin$^{-2}$. The peak is centred in the proximity of MRSS252-054422 (galaxy A), whilst we note that a potentially filamentary structure extends $\approx3-4'$ to the South-East. Being outside the MUSE field of view, we have no way of distinguishing whether this extension consists of $z=0.28$ galaxies or is simply a projection effect.

\begin{figure}
	\includegraphics[width=\columnwidth]{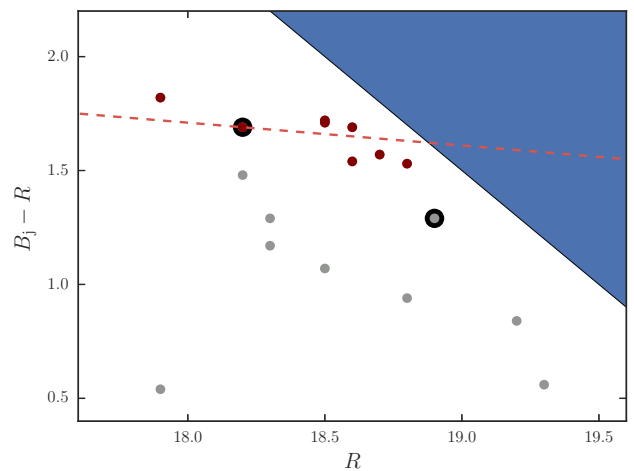}
    \caption{$(B_{\rm j}-R)$ versus $R$ colour-magnitude data points (small circles) for all galaxies detected in both MRSS $R$ band and APM $B_{\rm j}$ band within $5'$ of the QSO HE0515-4414. The two ringed points denote galaxies identified at $z=0.28$ from our MUSE data (i.e. galaxy A at $R=18.2$ and galaxy C at $R=18.9$). The dashed line shows a model red-sequence based on the $(B_{\rm j}-R)$ colour of galaxy A (MRSS~252-054422). Galaxies conforming with this model within $\pm0.2$~mag are shown in red and are highlighted by square boxes in Fig.~\ref{fig:ancillary}. The shaded region shows the limit of the APM $B_{\rm j}$ data.}
    \label{fig:redseq}
\end{figure}

As a further check, in Fig.~\ref{fig:redseq}, we show the $(B_{\rm j}-R)$ versus $R$ colour-magnitude diagram for galaxies detected by both the APM Galaxy Survey and MRSS (small circles). Galaxy A (non-SF, $R=18.2$~mag) and MRSS~252-053745 (galaxy C, SF, $R=18.9$~mag) from our observations are both detected in the APM Galaxy Survey and the MRSS, and are each highlighted with a large circle symbol. MRSS~252-054266 (galaxy B, $R=19.1$~mag) is not detected in the $B_{\rm j}\leq20.5$ complete galaxy sample form of the APM Galaxy Survey used here. The shaded region in Fig.~\ref{fig:redseq} shows the area of the colour-magnitude space lying beyond the $B_{\rm j}=20.5$ magnitude limit.

At $z=0.28$, the 4000~\AA\ break falls directly between these two bands, giving rise to a large $(B_{\rm j}-R)$ colour and making these bands ideal for detecting passive galaxies. Based on the $(B_{\rm j}-R)$ colour of galaxy A, we show a $z=0.28$ $(B_{\rm j}-R)$ versus $R$ red sequence model \citep[e.g.][]{2010A&A...523A..66B}, based on which we identify candidate $z=0.28$ passive galaxies surrounding the MUSE observations (we note that this model is consistent with the passive galaxy B not being detected in the APM $B_{\rm j}\leq20.5$ galaxy sample). Galaxies lying within $\Delta(B_{\rm j}-R)\leq0.2$ of the model red-sequence are denoted by red circles and are marked in Fig.~\ref{fig:ancillary} by squares. The candidate red-sequence galaxies are widely spread over scales of $\sim4'$ ($\sim1.3$~Mpc at $z=0.28$) and do not form any visible clump consistent with a significantly larger structure around the $z=0.28$ galaxies. 

Based on the 5 group members with MUSE redshifts, we estimate the group redshift and velocity dispersion to be $z_{\rm group}=0.2825\pm0.0002$ and $\sigma_{\rm r}=114^{+49}_{-35}$~km~s$^{-1}$ respectively \citep{1980A&A....82..322D}. Taking this velocity dispersion and assuming the group to be virialised results in a mass estimate of $M_h=6.4^{+4.2}_{-4.1}\times10^{12}$~M$_\odot$, with an estimated virial radius of $R_h=340\pm210$~kpc \citep[comoving,][]{1989ApJ...344...57R}. These estimates are consistent with the observed galaxies being part of a relatively low mass group \citep[e.g.][]{1989ApJ...344...57R,1998ApJ...496...39Z}, although given the limited view of this group of galaxies and the small number of member galaxies observed, the estimated virial mass and radius represent lower-limits (assuming the structure is in fact virialised).

Combining the above estimates with the lack of a cluster-scale structure in Fig.~\ref{fig:ancillary}, the MUSE observations and the QSO sightline appear to be probing the central regions of a galaxy group at $z=0.2825$ and we can reasonably exclude the possibility of this being the outskirts (or the core) of a galaxy cluster. We note that this is not absolutely conclusive however in terms of the large scale structure surrounding the MUSE field of view. Planned integral field unit (IFU) spectroscopy observations with the K-band Multi Object Spectrograph (KMOS) on the Very Large Telscope (VLT) and grism observations with Wide Field Camera 3 (WFC3) on the Hubble Space Telscope (HST) will clarify this conclusion.

\subsection{Group galaxy properties}

\begin{table*}
\caption{Group galaxy properties.}
\label{tab:galprops}
\begin{tabular}{lcccccccccc}
\hline
ID             & & $b$   & $f($H~$\alpha)$             & SFR              & \multicolumn{2}{c}{P.A.}      & \multicolumn{2}{c}{$V_{\rm max}$} & \multicolumn{2}{c}{$\sigma_{\rm v}$}\\                
               & & (kpc)& ($10^{-18}$erg/s/cm$^{2}$) & (M$_{\odot}$~yr$^{-1}$) & \multicolumn{2}{c}{($^\circ$)}& \multicolumn{2}{c}{(km~s$^{-1}$)} & \multicolumn{2}{c}{(km~s$^{-1}$)} \\
               & &  &                         &                  & Gas        & Stellar     & Gas       & Stellar    & Gas        & Stellar \\
\hline
MRSS~252-054422 & (A) & 132 & $54.7\pm2.8$                & $0.169\pm0.009$  & ---        & $159\pm7$   & ---       & $135\pm32$ & ---        & $158\pm7$ \\
MRSS~252-054266 & (B) &  88 & $<5.0$                      & $<0.015$         & ---        & $102\pm10$  & ---       & $85\pm23$  & ---        & $160\pm10$ \\
MRSS~252-053745 & (C) & 101 & $1077\pm11$                 & $3.236\pm0.032$  & $35\pm8$   & $51\pm55$   & $23\pm5$  & $32\pm37$  & $34\pm2$   & $91\pm22$ \\
J051708-441039  & (D) &  83 & $314.2\pm9.4$               & $0.943\pm0.028$  & $116\pm22$ & ---         & $38\pm12$ & ---        & $68\pm5$   & --- \\
J051708-441103  & (E) &  49  & $68.4\pm3.2$                & $0.208\pm0.010$  & ---        & ---         & ---       & ---        & ---        & --- \\
\hline
\end{tabular}
\end{table*}

For each of the member group galaxies, we calculate the total H~$\alpha$ flux and from this estimate the integrated star-formation rate (SFR). Assuming a Kroupa 2003 IMF \citep{2003ApJ...598.1076K}, the SFRs are estimated with $C_{\rm H\alpha}=5.5\times10^{42}$~M$_{\odot}$~yr$^{-1}$~ergs$^{-1}$~s \citep{2011ApJ...737...67M,2011ApJ...741..124H}, whereby ${\rm SFR}_{\rm H\alpha}=C_{\rm H\alpha}L_{\rm H\alpha}$. We then correct the SFR estimates for intrinsic absorption assuming $A_{\rm V}=1.0$. The resulting corrected SFRs are presented in Table~\ref{tab:galprops}, alongside the impact parameter (given in proper coordinates) and the galaxy dynamical information derived in the sections that follow. For the two `passive' galaxies in the group we place an upper limit on the SFR of $<0.015$~M$_{\odot}$~yr$^{-1}$ for MRSS~252-054266, whilst MRSS~252-054422 (i.e. our assumed brightest group member) shows low levels of star formation based on the detection of relatively weak H~$\alpha$ emission. 

\subsection{Absorption features at the group redshift}
\label{sec:absfeat}

From the optical CASPEC data of HE0515-4414, \citet{2000A&A...363...69D} found absorption lines for Al~{\sc i}, Ca~{\sc ii}, Fe~{\sc i}, Fe~{\sc ii}, Mn~{\sc ii} and Mg~{\sc ii} at $z=0.281-0.282$. Based on their Mn~{\sc ii} column density and assuming Mn abundance relative to H of 1/10 solar, \citet{2000A&A...363...69D} estimate a H~{\sc i} column density of $N($H{\sc i}$)\approx10^{19.9}$~cm$^{-2}$.

\begin{figure}
	\includegraphics[width=\columnwidth]{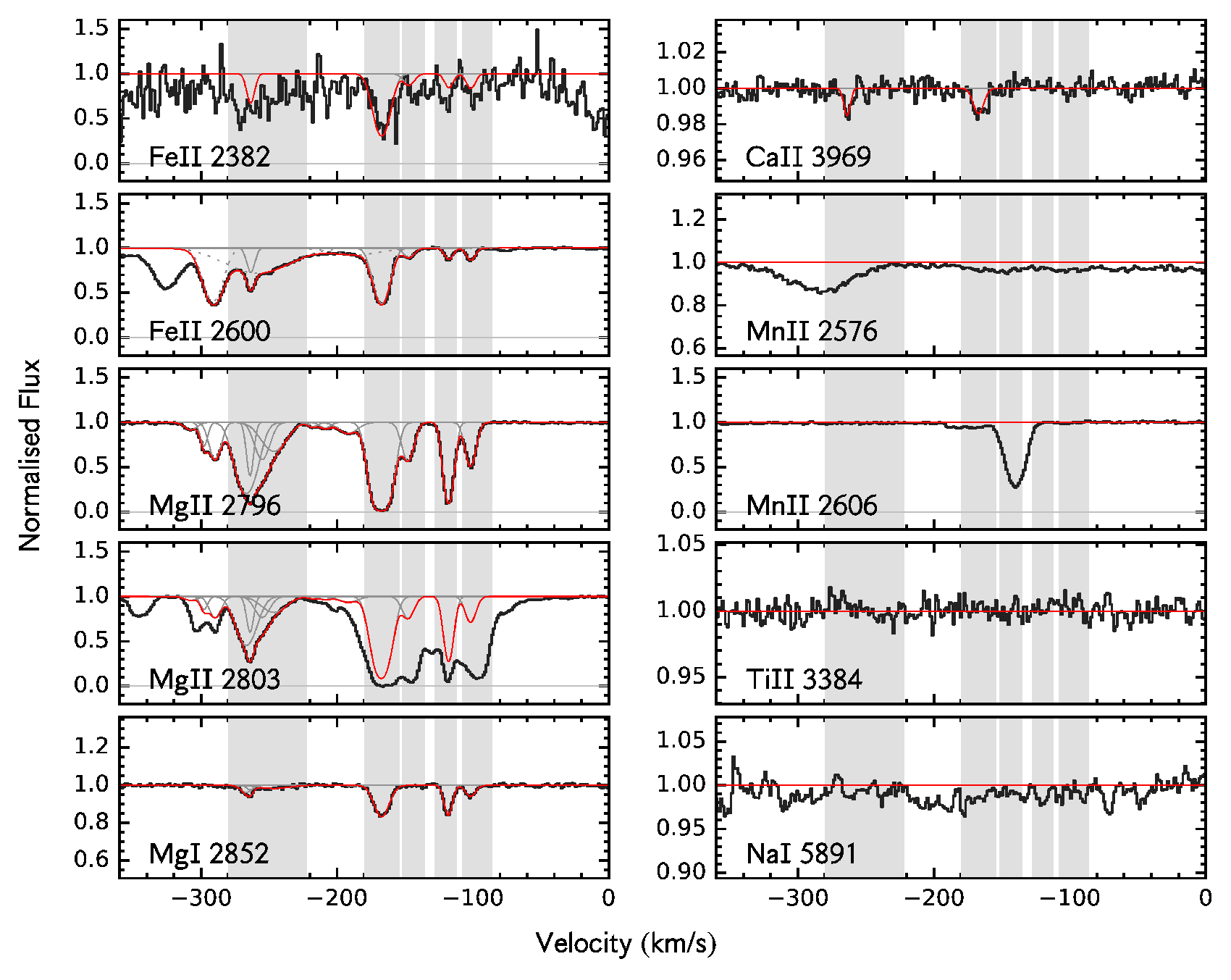}
    \caption{The UVES spectra (black line) and fitted profiles (red line) for the transitions given in Table~\ref{tab:UVESabsorbers}. The velocity scale along the horizontal axis gives the offset from the estimated group centre (where galaxies C and D lie at velocities of $\approx-50-100$~km~s$^{-1}$ and galaxies A, B and E lie at $\gtrsim0$~km~s$^{-1}$). Individual Voigt profile components are shown by the grey curves. The five shaded grey areas in each plot show the regions used to estimate the column densities given in Table~\ref{tab:UVESabsorbers} (we do not attempt to estimate column densities for lines outside these regions due to blending issues).}
    \label{fig:uves_fitting}
\end{figure}

The metal lines listed above are all covered by available UVES spectra, which we have used to re-analyse the absorber properties. There are five components for which we measure individual column densities and $1\sigma$ errors using the apparent optical depth (AOD) method. We note that Fe~{\sc ii} is blended with forest absorption, so $N_{\rm FeII}$ is instead estimated from Voigt profile (VP) fitting. The column densities, the values of $\lambda_{\rm min}$ and $\lambda_{\rm max}$, and the velocities relative to $z=0.2825$ are given in Table~\ref{tab:UVESabsorbers}. The minimum ($\lambda_{\rm min}$) and maximum ($\lambda_{\rm max}$) wavelengths indicate the ranges used for the above measurements. The estimated errors on the column densities include zero-level and continuum uncertainties. We plot the continuum  normalised flux profiles from the UVES data for the available Fe~{\sc ii}, Mn~{\sc ii} and Mg~{\sc ii} lines in Fig.~\ref{fig:uves_fitting} (black lines). Fig.~\ref{fig:uves_fitting} also shows the total VP fit (red curve), the individual component VP fits (grey curves - where not obscured by the total VP fit), and the regions used to estimate the column densities given in Table~\ref{tab:UVESabsorbers} (grey shaded areas). From the UVES data, we find that the total rest equivalent width of the Mg~{\sc ii} absorption is $W_{\rm r}(2796)=0.73\pm0.02$~\AA.



\begin{table*}
\caption{Properties of the absorbers detected in the UVES spectrum close to the galaxy redshifts. The upper and lower limits on a number of the ion column densities are given at the $3\sigma$ level.}
\label{tab:UVESabsorbers}
\begin{tabular}{rlcccccccc}
\hline
Atom & Ion & $\lambda_{\rm rest}$ & $v_{\rm min}$ & $v_{\rm max}$ & $\lambda_{\rm min}$ & $\lambda_{\rm max}$ & $\log N$ \\
 &  & (\AA) & (km\,s$^{-1}$) & (km\,s$^{-1}$) & (\AA) & (\AA) & (cm$^{-2}$) \\
\hline
Ca & II & $3969.590$ & $-272.0$ & $-255.0$ & $5086.383$ & $5086.670$ & $10.55^{+0.19}_{-0.30}$ \\
Fe & II & $2600.172$ &   $-$    &   $-$    &   $-$    &   $-$    & $12.16\pm0.05$ \\
Mg & I & $2852.963$ & $-273.8$ & $-230.9$ & $3655.585$ & $3656.108$ & $10.85^{+0.13}_{-0.17}$ \\
Mg & II & $2796.354$ & $-316.9$ & $-224.6$ & $3582.535$ & $3583.638$ & $>13.08$ \\
Mn & II & $2606.462$ & $-279.1$ & $-221.9$ & $3339.677$ & $3340.314$ & $<11.86$ \\
Na & I & $5891.583$ & $-280.0$ & $-226.7$ & $7548.903$ & $7550.245$ & $<10.77$ \\
Ti & II & $3384.730$ & $-281.5$ & $-236.1$ & $4336.842$ & $4337.500$ & $<10.99$ \\
\hline
Ca & II & $3969.590$ & $-175.8$ & $-156.3$ & $5088.015$ & $5088.346$ & $10.71^{+0.17}_{-0.19}$ \\
Fe & II & $2600.172$ & $-180.1$ & $-154.1$ & $3332.719$ & $3333.008$ & $12.94^{+0.03}_{-0.02}$ \\
Mg & I & $2852.963$ & $-180.2$ & $-152.9$ & $3656.727$ & $3657.060$ & $11.25^{+0.04}_{-0.04}$ \\
Mg & II & $2796.354$ & $-184.3$ & $-154.5$ & $3584.120$ & $3584.477$ & $>13.12$ \\
Mn & II & $2576.877$ & $-178.3$ & $-153.6$ & $3302.880$ & $3303.152$ & $<11.65$ \\
Na & I & $5891.583$ & $-181.2$ & $-155.2$ & $7551.391$ & $7552.046$ & $<10.74$ \\
Ti & II & $3384.730$ & $-180.1$ & $-155.5$ & $4338.309$ & $4338.666$ & $<10.74$ \\
\hline
Ca & II & $3969.590$ & $-152.4$ & $-134.2$ & $5088.412$ & $5088.721$ & $<10.32$ \\
Fe & II & $2600.172$ & $-154.1$ & $-137.1$ & $3333.008$ & $3333.196$ & $11.83^{+0.04}_{-0.04}$ \\
Mg & I & $2852.963$ & $-151.7$ & $-133.5$ & $3657.075$ & $3657.297$ & $<10.29$ \\
Mg & II & $2796.354$ & $-154.5$ & $-136.2$ & $3584.477$ & $3584.695$ & $12.17^{+0.03}_{-0.02}$ \\
Mn & II & $2576.877$ & $-154.9$ & $-136.7$ & $3303.138$ & $3303.338$ & $<11.59$ \\
Na & I & $5891.583$ & $-152.5$ & $-138.2$ & $7552.112$ & $7552.472$ & $<10.31$ \\
Ti & II & $3384.730$ & $-152.9$ & $-136.0$ & $4338.704$ & $4338.948$ & $<10.71$ \\
\hline
Ca & II & $3969.590$ & $-130.2$ & $-105.6$ & $5088.788$ & $5089.207$ & $<10.41$ \\
Fe & II & $2600.172$ & $-125.5$ & $-108.5$ & $3333.326$ & $3333.514$ & $11.83^{+0.05}_{-0.05}$ \\
Mg & I & $2852.963$ & $-124.4$ & $-110.0$ & $3657.408$ & $3657.583$ & $10.97^{+0.04}_{-0.04}$ \\
Mg & II & $2796.354$ & $-128.5$ & $-109.0$ & $3584.788$ & $3585.021$ & $>12.59$ \\
Mn & II & $2576.877$ & $-125.0$ & $-111.9$ & $3303.467$ & $3303.611$ & $<11.33$ \\
Na & I & $5891.583$ & $-131.7$ & $-109.7$ & $7552.636$ & $7553.192$ & $<10.56$ \\
Ti & II & $3384.730$ & $-132.1$ & $-111.3$ & $4339.005$ & $4339.306$ & $<10.63$ \\
\hline
Ca & II & $3969.590$ & $-112.0$ & $-88.7$ & $5089.097$ & $5089.494$ & $<10.54$ \\
Fe & II & $2600.172$ & $-108.5$ & $-93.0$ & $3333.514$ & $3333.687$ & $11.85^{+0.04}_{-0.04}$ \\
Mg & I & $2852.963$ & $-110.0$ & $-91.8$ & $3657.583$ & $3657.805$ & $10.61^{+0.10}_{-0.12}$ \\
Mg & II & $2796.354$ & $-109.0$ & $-89.5$ & $3585.021$ & $3585.254$ & $12.14^{+0.03}_{-0.02}$ \\
Mn & II & $2576.877$ & $-111.9$ & $-97.7$ & $3303.611$ & $3303.768$ & $<11.46$ \\
Na & I & $5891.583$ & $-109.7$ & $-86.3$ & $7553.192$ & $7553.782$ & $<10.55$ \\
Ti & II & $3384.730$ & $-111.3$ & $-89.2$ & $4339.306$ & $4339.626$ & $<10.51$ \\
\hline
\end{tabular}
\end{table*}

At the group redshift, the STIS E230M spectra cover a rest-frame wavelength range of $\lambda\approx1770-2430$~\AA, providing coverage of the common transitions of Si~{\sc ii}, Fe~{\sc ii}, and Al~{\sc iii}. Indeed we find multiple Fe~{\sc ii} absorption lines at $z=0.28138$ and $z=0.28177$, plus Si~{\sc ii}~1808~\AA\ at $z=0.28135$.

\subsection{Galaxy velocity fields}

The co-incidence of the absorbers and group galaxies may be the result of a range of possible connections. The absorbers may be tracing gas that could be physically attributed (i.e. bound) to the halo of a single galaxy; they could be probing gas within a shared intra-group medium and not bound to any single galaxy; or they may be attributable to outflowing gas clouds, driven by ongoing or recent star-formation. To help disentangle these possibilities, it is useful to derive information on the dynamics of both gas and stars within the galaxy disks in the proximity of the absorbing gas.

The MUSE data presented here offer the opportunity to do exactly that and in this section we present the results of analyses of both the gas and stellar dynamics of the group galaxies.

\subsubsection{Gas dynamics}

We measure the gas dynamics for the group member galaxies using the observed H~$\alpha$ emission. This is observed in 4 of the detected galaxies, however J051708-441103 (galaxy E) is not resolved sufficiently (given the measured half-light radius of $R_{e}=0.52$, consistent with a point source under the seeing conditions) and the H~$\alpha$ emission detected in galaxy A does not possess a high enough SNR to enable a meaningful 2D velocity analysis to be made. As such we present the gas velocity fields for galaxy C (MRSS~252-053745) and galaxy D (J051708-441039).

\begin{figure}
	\includegraphics[width=\columnwidth]{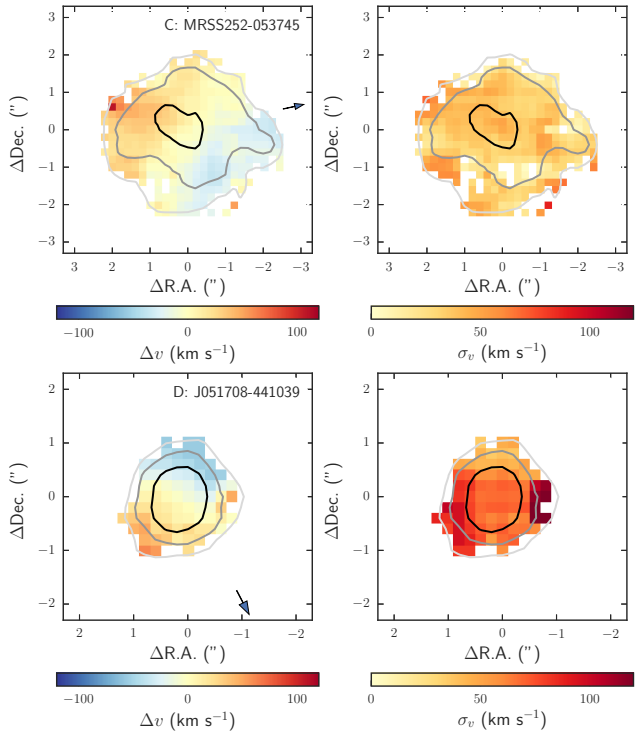}
    \caption{The H~$\alpha$ emission velocity field (left column) and velocity dispersion (right column) of the two brightest H~$\alpha$ emitters in our sample: galaxy C (MRSS~252-053745, top); and galaxy D (J051708$+$441039, bottom). In each case the colour-map shows the velocity measurement, whilst the contours show the $3\sigma$, $6\sigma$, and $9\sigma$ S/N levels from the collapsed H~$\alpha$ emission image. The direction to the QSO sightline is given by the arrow in each of the left hand panels.}
    \label{fig:gvelfit}
\end{figure}

\begin{figure}
	\includegraphics[width=\columnwidth]{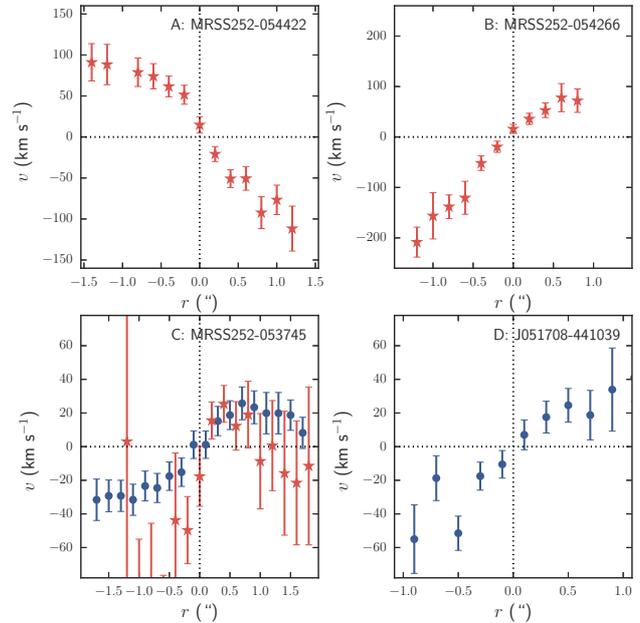}
    \caption{Rotation curves derived from the gas dynamics (traced via H~$\alpha$ emission, blue circles) and stellar dynamics (traced by SSPs, red stars).}
    \label{fig:rotcurves}
\end{figure}

For each galaxy, we fit the H~$\alpha$ emission in all spatial IFU pixels with an integrated H~$\alpha$ flux of SNR>2. The fit is performed with a Gaussian function, from which we calculate the velocity offset ($\Delta v$) and line-of-sight velocity dispersion ($\sigma_{\rm v}$) in each pixel. The resulting maps are shown in Fig.~\ref{fig:gvelfit}, with $\Delta v$ shown by the colour maps in the left hand column and $\sigma_{\rm v}$ shown by the colour maps in the right hand column. In both columns, the contours show the 2$\sigma$, 4$\sigma$, and 8$\sigma$ SNR levels of the integrated H~$\alpha$ flux. From these maps, we estimate the position angles for the two galaxies using {\sc fit\_kinematic\_pa} \citep{2006MNRAS.366..787K}. Using the derived position angles, we then calculate rotation curves for each galaxy, which are shown by the circular points in the lower two panels of Fig.~\ref{fig:rotcurves}.

The calculated positional angles, maximal velocity offsets ($V_{\rm max}$) and mean velocity dispersions ($\sigma_{\rm v}$) derived from the gas kinematics for these two galaxies are listed in Table~\ref{tab:galprops}. The estimated inclination angles for galaxies C and D are $i=24.9\pm3.2$ and $i=75.7\pm10.1$ respectively.

\subsubsection{Stellar dynamics}

We measure the galaxy velocity fields for our three highest S/N galaxies (i.e. galaxies A, B and C) using the {\sc ppxf} python code \citep{2004PASP..116..138C}. For the template fitting, we use the \citet{1999ApJ...513..224V} Single Stellar Population (SSP) library, fitting to our observed spectra over a rest-frame wavelength range of $4800~\AA<\lambda<5465~\AA$.

\begin{figure}
	\includegraphics[width=\columnwidth]{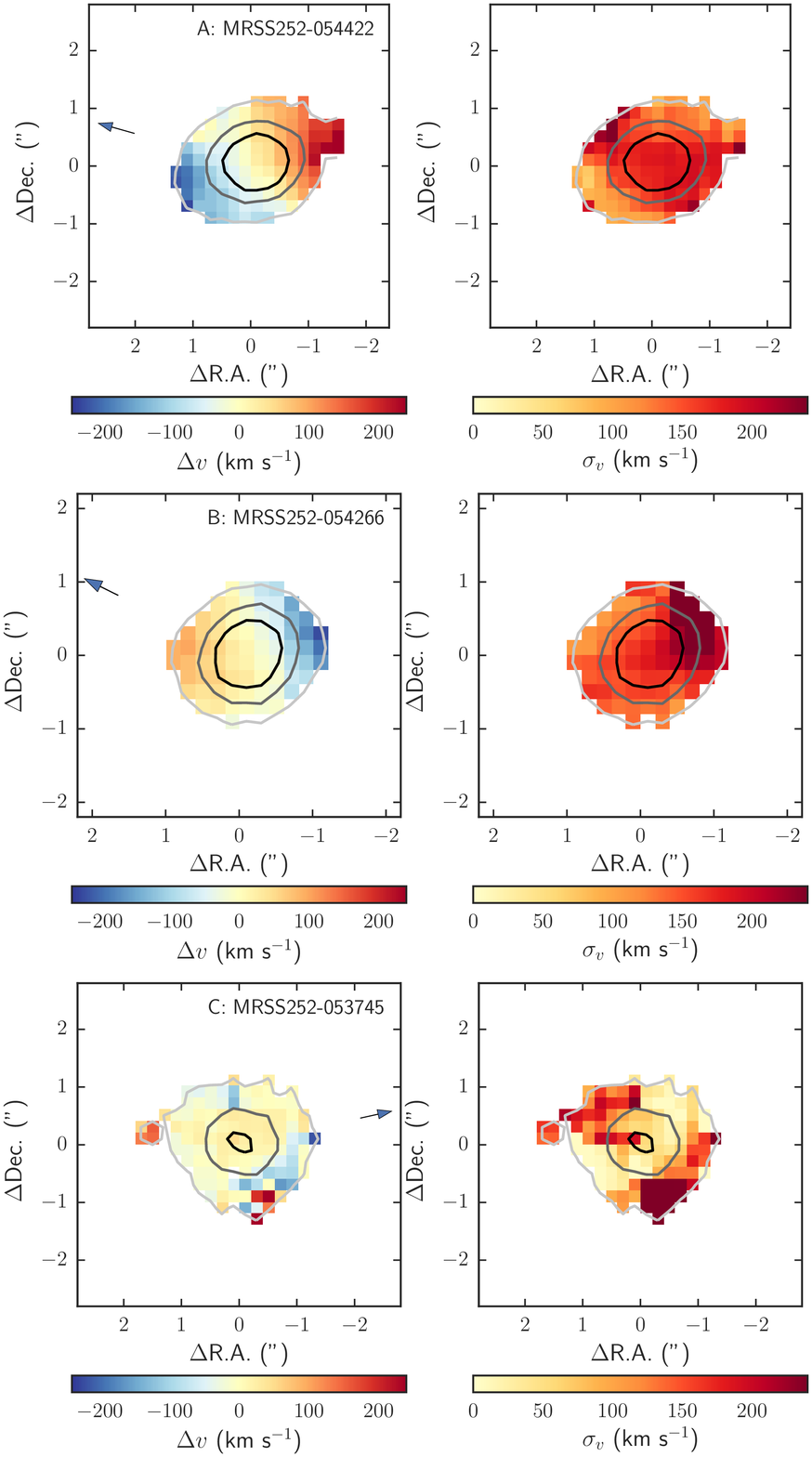}
    \caption{The stellar absorption velocity field (left column) and velocity dispersion (right column) of the three brightest and most extended $z=0.28$ galaxies: galaxy B (MRSS~252-054266, top); galaxy A (MRSS~252-054422, middle); and galaxy C (MRSS~252-053745, bottom). In each case the colour-map shows the velocity measurement, whilst the contours show the $3\sigma$, $6\sigma$, and $9\sigma$ S/N levels from the collapsed continuum image (over the wavelength range covered in the fitting procedure). The direction to the QSO sightline is given by the arrow in each of the left hand panels.}
    \label{fig:velfit}
\end{figure}

The results of the template velocity fitting with {\sc ppxf} are shown in Fig.~\ref{fig:velfit}, where the left column shows the velocity field for each galaxy and the right column shows the velocity dispersion. In each of the velocity field panels, the direction to the QSO sightline is shown by the arrow. In all panels, the contour curves show the $3\sigma$ (pale grey), $6\sigma$ (dark grey) and $9\sigma$ (black) S/N levels from the collapsed continuum image. Again the calculated positional angles, maximal velocity offsets ($V_{\rm max}$) and mean velocity dispersions ($\sigma_{\rm v}$) are given in Table~\ref{tab:galprops}.

Clear rotational dynamics are seen in the stellar population for both of the passive galaxies (i.e. galaxies A and B). We note that the sides of the two galaxies nearest to each other are both blueshifted, such that the two galaxies are rotating in the opposite direction to each other. Again the rotation curves for these galaxies are shown in Fig.~\ref{fig:rotcurves} (red stars in the top two panels), showing the clear rotational dynamics detected in the MUSE data.

The rotation in the stellar component of galaxy C is less clear, although we note that the blueshifted component in the South-West of the galaxy is consistent with the blueshifted component seen in the gas dynamics in the previous section. This is evident in the rotation curve shown in Fig.~\ref{fig:rotcurves}, where the red stars show the stellar rotation curve and the blue circles show the gas dynamics. The stellar rotation curve is much noisier than the gas dynamics, but does at least show agreement in terms of polarity of the rotation.

\section{Discussion}
\label{sec:discussion}

The MUSE data on this group/QSO sightline combination provides a wealth of information on the group members. In particular the galaxy velocity maps provide an opportunity to disentangle the relationship between the absorbers and the galaxies. We now explore a number of possibilities for what the structure is that the absorbing gas is tracing: the co-rotating halo of one of the group galaxies; an outflow from one of the group galaxies; and material within a wider group medium.

Before considering the possibilities associated with the absorber probing halo gas of one of the absorbed galaxies, we first need to evaluate the possibility of there being an undetected galaxy, either hidden by the QSO light or below the detection threshold of the data, that the absorbers could instead be associated with.

To evaluate the potential for a star-forming galaxy directly in the line of sight of the QSO, we perform a subtraction of the QSO continuum (incorporating the point spread function on the QSO) within $\pm1000$~km~s$^{-1}$ of the galaxy group redshift. We find no evidence for any line emission in this window, placing an upper limit on any H~$\alpha$ emission at $z=0.2825$ coincident with the QSO sightline of $f({\rm H\alpha}) \leq 8.4 \times 10^{-18}$~erg~cm$^{-2}$~s$^{-1}$~\AA$^{-1}$. We note also that should there be a coincident system in the line of sight, statistically we might expect a much larger Mg~{\sc ii} absorber equivalent width than observed.

\subsection{Co-rotating disk/halo}

Perhaps the simplest physical model for the nature of the sightline absobers in relation to the galaxy population is that they inhabit a gaseous halo co-rotating with a single host galaxy \citep[e.g.][]{2002ApJ...570..526S,2005pgqa.conf...24C}. In this simple model, it may be expected that the absorbers possess a velocity offset consistent with extrapolating the velocity curve of the host galaxy out to the distance of the absorbers \citep[e.g.][]{2001AJ....122..679C,2005pgqa.conf...24C,2002ApJ...570..526S}.

Such a configuration is predicted to result from infall of material from the IGM onto galaxy halos \citep[e.g.][]{2011ApJ...738...39S,2013ApJ...769...74S} and so cool absorber species such as Mg~{\sc ii} are predicted to be tracers of infalling material \citep[e.g.][]{2014PASP..126..969B}. Recent observations have shown evidence for such co-rotating material potentially resulting from infall onto galaxy halos with for example \citet{2016ApJ...824...24D} and \citet{2016ApJ...820..121B} presenting absorption line studies in the proximity of a star-forming galaxies at $z\sim0.4$ and another at $z\sim1$ respectively. Such co-rotating halo detections have exclusively been claimed for isolated galaxies thus far. Although some potential detections of infall onto groups has been presented \citep{2011ApJ...743...95G,2014A&A...570A..16C}, these observations are at high redshift ($z\sim2-3$) and are posited as evidence for filamentary (i.e. cold-flow) accretion \citep[which is predicted to dominate gas accretion processes at $z\gtrsim2$, e.g.][]{2009Natur.457..451D,2011MNRAS.415.2782V,2011MNRAS.413L..51K,2011MNRAS.418.1796F,2011MNRAS.413L..51K,2012MNRAS.424.2292G}.

The most closely associated galaxy in our sample to the Mg~{\sc ii} absorbers, in terms of on-sky separation, is galaxy E, with $b=49$~kpc and $M_{\rm B}=-18.13$. It is also the most separated in velocity space with $\Delta v\approx300-500$~km~s$^{-1}$. Galaxy E has no velocity information from the MUSE data given its relatively low surface brightness. \citet{2010ApJ...714.1521C} investigate what they refer to as extended Mg~{\sc ii} absorbing halos using a large sample of Mg~{\sc ii} absorber-galaxy pairs. They plot the Mg~{\sc ii} equivalent width, $W_{\rm r}(2796)$, of absorption lines as a function of a combination of nearest galaxy distance, $b$, and absolute magnitude, $M_{\rm B}$, finding evidence of a correlation. Using the impact parameter and brightness of galaxy E with the relation derived by \citet{2010ApJ...714.1521C} (measured for Mg~{\sc ii} absorbers at a median redshift of $\left<z\right>=0.2357$) predicts a Mg~{\sc ii} equivalent width of ${\rm log}W_{\rm r}(2796)=-1.04\pm0.37$ (assuming $M_{\rm B}^\star=-20.3$, \citealt{2007ApJ...665..265F}). This falls significantly below the observed total rest equivalent width of the Mg~{\sc ii} absorption (${\rm log}W_{\rm r}(2796)=-0.137\pm0.012$), with the observation lying above the relation by $2.7\sigma$ \citep{2010ApJ...714.1521C}. 

Galaxy E would seem an unusual Mg~{\sc ii} `host' galaxy then, given its outlier status in the relationship presented by \citet{2010ApJ...714.1521C}. Given the close proximity of the four further $z=0.28$ galaxies, and the relatively large velocity offset between galaxy E and the absorber system, we now consider the dynamics of the four brighter galaxies detected close to the sightline. 

Galaxy D has the smallest offset in velocity space to the absorber group and lies at only $b=83$~kpc from the sightline. From the ISM gas kinematics, we find that the absorbers are on the redshifted side of the galaxy, however they lie blueshifted from the galaxy redshift ($-10\lesssim\Delta v\lesssim-200$~km~s$^{-1}$). They are thus offset in velocity in the opposite direction to the galaxy disk rotation.

The remaining three galaxies lie at impact parameters of $\gtrsim100$~kpc from the sightline, which based on the correlation of \citet{2010ApJ...714.1521C} implies that these galaxies are unlikely to host a ${\rm log}W_{\rm r}(2796)=-0.137$ absorber in such a discrete single halo model. For example, an equivalent width of ${\rm log}W_{\rm r}(2796)=-0.60\pm0.34$ is predicted based on the impact parameter and luminosity of galaxy C, significantly lower than observed.

Briefly considering the dynamics of these three galaxies however, we note that: for B the absorbers lie $\approx-100$~km~s$^{-1}$ (i.e. bluewards) of the galaxy, but are closest to the redshifted side of the galaxy; for C the sightline is aligned with the blueshifted ($\Delta v\approx-30$~km~s$^{-1}$) side of the galaxy with the absorber components covering a velocity range of $-30\lesssim\Delta v\lesssim-220$~km~s$^{-1}$ relative to the intrinsic galaxy redshift; and for A the absorbers are offset by $\Delta v\gtrsim-100$~km~s$^{-1}$ and lie on the blueshifted side ($\approx-120$~km s$^{-1}$) of the galaxy. The absorbers are thus aligned in velocity space with both the rotational kinematics of galaxies C and A, but anti-aligned for galaxy B.

Although we find kinematic alignment of the absorbers with the galactic rotation in the cases of both galaxy A and galaxy C, the high equivalent width makes such a scenario unlikely. In addition, given the comparable impact parameters (101~kpc versus 132~kpc) of these two galaxies, it is difficult to justify a gravitational association between the absorption system and just one of the galaxies. We conclude then that the observations do not favour a single galaxy-absorber pair with a co-rotational configuration.

In summary, such a disk co-rotation model is an unlikely configuration for this system given the data available.


\begin{figure*}
	\includegraphics[width=15cm]{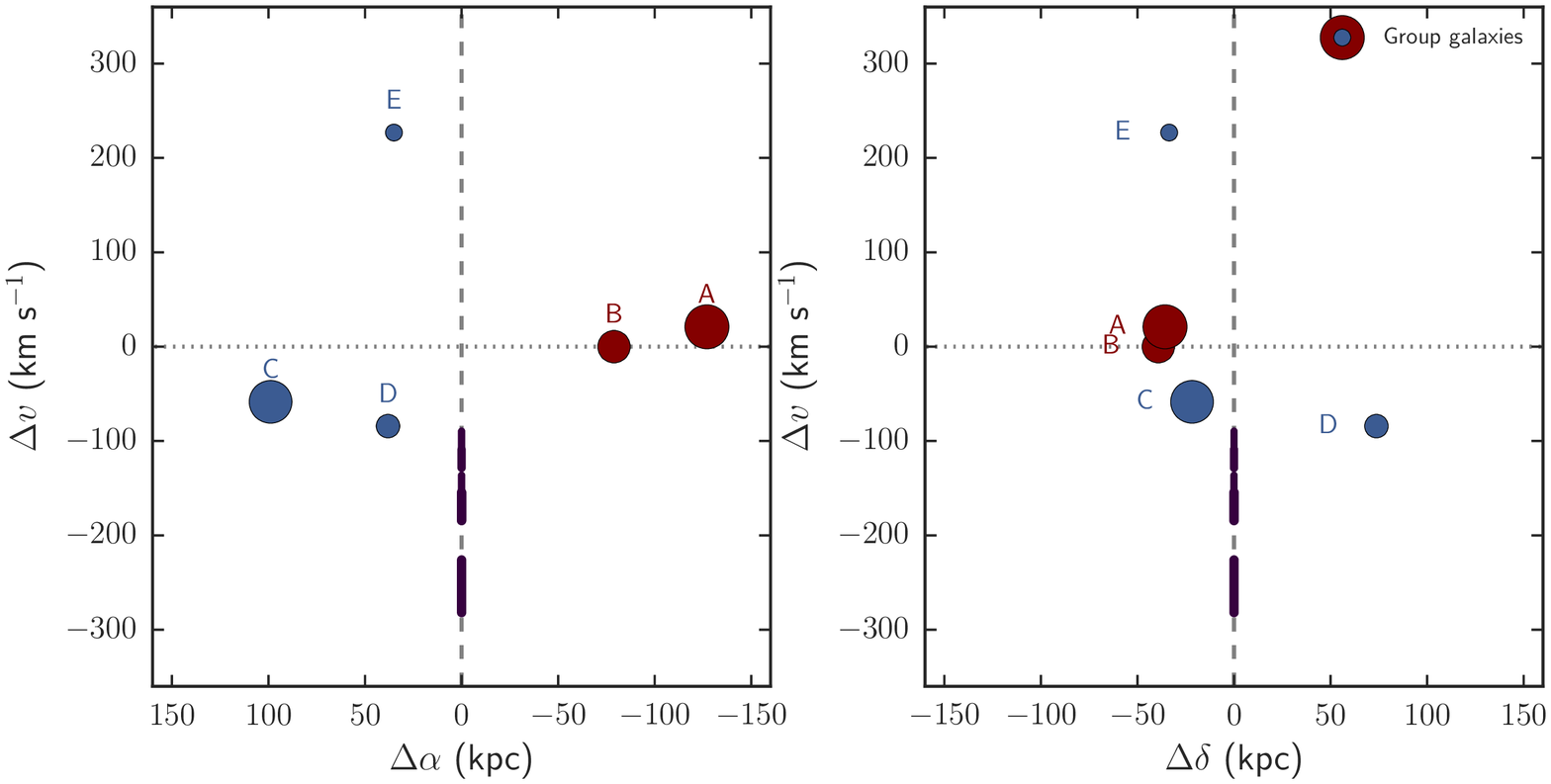}
	\includegraphics[width=12cm]{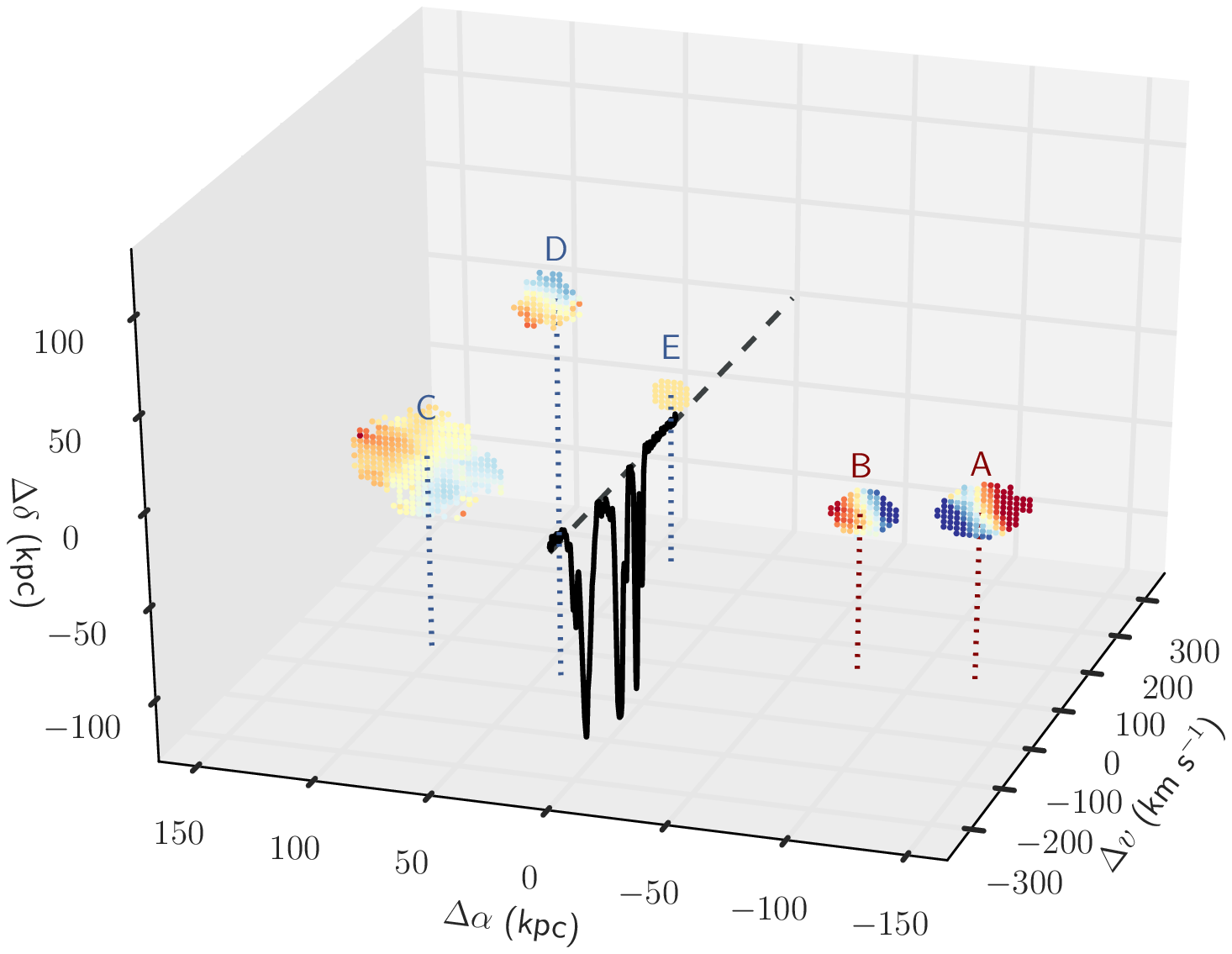}
    \caption{\emph{Top panels}: The distribution of group galaxies (circles) around the QSO sightline in velocity space and impact parameter projected in R.A. (left) and Declination (right). Passive galaxies are shown in red and star-forming galaxies are shown in blue, whilst the size of points is proportional to the measured galaxy half light radius. The dashed line indicates the QSO sightline, whilst the solid sections along this line show the velocity space covered by the Mg~{\sc ii} absorbers listed in Table~\ref{tab:UVESabsorbers}. \emph{Lower panels}: The same galaxies shown in 3D projection around the sightline. For galaxies A, B, C and D, the measured galactic kinematics are shown (magnified by a factor of 3 in $\Delta\alpha$ and $\Delta\delta$) by the colour gradients. Dotted vertical lines are plotted to give a sense of the distribution, with the lower point of each line positioned on the $\Delta\alpha-\Delta v$ axis plane. The dashed line shows the position of the QSO sightline, whilst we show the normalised UVES spectrum (scaled in velocity space to Mg~{\sc ii}~2796~\AA) for illustration purposes.}
    \label{fig:3dimpact}
\end{figure*}

\subsection{Outflows}

Alternatively, there is a broad range of work suggesting that Mg~{\sc ii} absorbers may be evidence of outflowing material \citep[e.g.][]{2011ApJ...743L..34K}. In the system observed here, this possibility is supported by the fact that the absorbing material is entirely blue-shifted from the observed galaxies. Absorber strength also provides some indication of whether the absorber system is consistent with tracing an outflow and it has been shown in systems where outflows are observed directly (and probed by the sightline of a background source) that they are almost exclusively associated with $W_r(2796)\gtrsim1$~\AA\ absorption systems. Considering the Mg~{\sc ii} system presented here, the equivalent width of $W_r(2796)=0.73\pm0.02$~\AA\ places it as a borderline candidate for outflowing material by this criterion.

The galaxy dynamical structure is often invoked in differentiating between the outflowing and infalling scenarios, primarily using the position angle as the defining parameter. The criteria used by  number of authors \citep[e.g.][]{2012MNRAS.426..801B, 2014ApJ...792L..12K, 2014ApJ...784..108B, 2014ApJ...794..130B, 2016ApJ...833...39S} applies the logic that if an absorber is aligned with the minor axis of a star-forming galaxy (i.e. the  azimuthal angle to the absorber is $\alpha\gtrsim45^\circ$) then it is assumed to be tracing outflowing material; whilst if an absorber is aligned with the major-axis of a star-forming galaxy (i.e. $\alpha\lesssim45^\circ$), it is assumed to be tracing infalling material. However, the results of \citet{2011ApJ...743...10B} show evidence for any azimuthal angle dependence of Mg~{\sc ii} absorber properties vanishing beyond impact parameters of $b\gtrsim50$~kpc.

All our detected galaxies lie at impact parameters beyond where any azimuthal angle dependency is observed. Additionally, the nearest detected galaxy (E) in our sample is not detected with sufficient signal-to-noise to afford a dynamical analysis anyway. We note that the absorption system does lie significantly blueshifted, at $\approx250-550$~km~s$^{-1}$, from galaxy E lending itself well to an outflow scenario for the absorber however. We measure a star-formation rate of $0.208\pm0.010$~M$_{\odot}$~yr$^{-1}$ for this source based on its measured H~$\alpha$ flux. Looking to statistical studies of outflows traced by Mg~{\sc ii}, \citet{2010ApJ...719.1503R} find in their sample that galaxies with SFR~$\lesssim4$~M$_{\odot}$~yr$^{-1}$ show associated Mg~{\sc ii} absorption equivalent widths of $W<0.44$~\AA (although they do measure Mg~{\sc ii} equivalent widths of $W\approx1$~\AA\ in galaxies with SFRs as low as $\approx1$~M$_{\odot}$~yr$^{-1}$). These SFRs are significantly higher than that observed for galaxy E and we conclude that, short of a previous strong star-burst episode, galaxy E is incapable of powering outflows giving rise to strong equivalent width absorbers.

The two most strongly star-forming galaxies in the sample, galaxies C ($b=101$~kpc) and D ($b=83$~kpc), lie at approximately the same velocity as the strongest of the Mg~{\sc ii} absorption, with the absorption complex extending $\sim100-150$~km~s$^{-1}$ bluewards of the galaxy velocities (see Fig.~\ref{fig:3dimpact}). With an SFR of 3.2~M$_{\odot}$~yr$^{-1}$, galaxy C would be the most likely of the two to power an outflow, however given the large impact parameter of $b\approx100$~kpc it is unlikely that the sightline absorber is close enough to be probing such an outflow.

With impact parameters of $b\gtrsim90$~kpc and low or negligible SFR, the two passive galaxies (i.e. A and B) appear unlikely sources for outflowing material. Finally, given the flux limits of our observations, it is unlikely that a galaxy below our detection threshold would be undergoing powerful enough star-formation to be powering such a wind.

In conclusion, if the absorber system is tracing outflowing gas it is most likely associated with galaxy C. This galaxy has the highest star-formation rate of our sample, whilst being the third nearest galaxy to the sightline given our detection limits. We note that one would be unlikely to conclude that this was outflowing gas based on assuming the absorber to be associated with the closest galaxy on the sky within $\approx\pm500$~km~s$^{-1}$, given galaxy E's low star-formation rate.

\subsection{Intra-group material}

From Fig.~\ref{fig:3dimpact}, it may seem unphysical to simply associate one or all of the sightline absorbers with a single galaxy within the group. All 5 galaxies found at $z=0.28$ lie within 200~kpc of the sightline, within the scales usually cited as typical for the CGM or halo of an $L^*$ galaxy \citep[e.g.][]{2013ApJ...777...59T}. Given the proximity of the galaxies to each other and to the position of the sightline between these galaxies, perhaps the most reasonable (and physical) explanation of the gas probed by the absorbers is as an intra-group medium consisting of the merged (or merging) halos of some combination of the identified galaxies. Indeed, in this case it seems more appropriate to consider the absorber impact parameter in terms of group-centre distance as opposed to galaxy distance and in this case we find $b_{\rm Gr}\approx110$~kpc assuming that the passive pair of galaxies forms the group centre. In terms of a group virial scale, this is equivalent to $\approx0.32R_{\rm h}$.

It is interesting to consider this system in the context of H~{\sc i} maps of the distribution of gas in the M81 group. \citet{1994Natur.372..530Y} presented the first high-resolution map of H~{\sc i} gas in the proximity of M81, M82 and NGC3077, showing a complex distribution of high column density gas bridges ranging within velocities of $\Delta v=\pm100$~km s$^{-1}$ of M81 and extending over on-sky distances of $\approx100$~kpc. The distribution of gas in the M81 group is highly indicative of tidally disrupted material that could be considered to be both an extension of the individual galaxies and the mixing of material to form an intra-group medium. The evidence for tidally disrupted material in the case presented here however is minimal, given the lack of any observed disruption to the groups galaxies.

The mixing of gas to form an intra-group medium could source material via two other possible scenarios: a) the gas is in the process of being accreted into the group halo, and b) the absorption represents multiple gas systems that are being accreted alongside their original parent galaxies to form the intra-group medium. Supporting the latter of these cases are the results and superposition model of \citet{2011ApJ...743...10B}. Comparing samples of nominally isolated galaxies and group galaxies, they find an enhancement in the total Mg~{\sc ii} equivalent width as measured within galaxy groups when compared to field galaxies. The authors model this with a superposition of the predicted equivalent widths for the individual group galaxies, finding that these linearly add up to the measured group galaxy equivalent widths. From Fig.~6 in \citet{2011ApJ...743...10B}, we find that at $b_{\rm Gr}\approx110$~kpc their average group model predicts $W_{\rm r}\approx0.2$~\AA. However, performing the superposition summation with the 5 detected group galaxies, we find a predicted equivalent width of $W_{\rm r}=0.79\pm0.07$~\AA, consistent with the measured value. It is highly plausible then that the absorption represents the projected overlap of halo gas from the individual galaxies and that the gas is being or has been accreted into the group environment alongside the group galaxies. This would suggest that the detected gas is at some stage in the process of forming the intra-group medium for this group.


Looking at comparable studies of cool gas in the proximity of galaxy group systems, \citet{2006MNRAS.368..341W} and \citet{2010MNRAS.406..445K} studied two separate groups of 5 galaxies (at $z=0.666$ and $z=0.313$ respectively). Both groups contain 4 galaxies within an impact parameter of 100~kpc of associated Mg~{\sc ii} absorbers. We derive a Mg~{\sc ii} equivalent width of $W_r(2796)=2.2\pm0.2$~\AA\ for the $z=0.666$ absorber based on UVES data of the background QSO reduced as part of the VLT LBG Redshift Survey \citep[VLRS,][]{2011MNRAS.414...28C}, whilst \citet{2010MNRAS.406..445K} quote $W_r(2796)=1.8$~\AA\ for the $z=0.313$ absorber.

\citet{2006MNRAS.368..341W} conclude that it is difficult to associate a given galaxy with the absorption system and that debris produced by interactions may be producing the absorption. \citet{2010MNRAS.406..445K} reach a similar conclusion, finding that 3 of the group galaxies exhibit perturbed morphologies and extended optical streams (extending up to $\sim25$~kpc) in HST data. This work further supports these previous studies, adding 2D dynamical data on an independent group-absorber system. Additionally, the present system is of a significantly lower Mg~{\sc ii} equivalent width, showing that such absorbers found associated with group systems may probe a wide range of absorber column densities. This likely reflects the wide range of substructure present in group systems and also in the halos of individual galaxies. Turning to the Local Group, we note that neither our own, nor the previous two works mentioned would have detected a significant fraction of the dwarf galaxy population observed in both the Milky Way and Andromeda halos \citep[e.g.][]{2016MNRAS.458L..59M,2016arXiv160503345C}. Such dwarf galaxy and tidal structure is observed to distances of $\gtrsim150$~kpc from both Milky Way and Andromeda \citep[e.g.][]{2009Natur.461...66M,2017ApJ...834..112K,2016arXiv160503345C}. Indeed, it is clear that `isolated' galaxies identified in the bulk of studies will also contain a vast range of substructure (in the form of undetected dwarf galaxies and tidally stripped material) which is undetected, hiding a complex halo environment.


\section{Summary and Conclusions}
\label{sec:conclusions}

We have presented MUSE observations of a galaxy group, which is itself intersected by a background QSO sightline containing multiple metal absorption lines coincident in velocity space with the galaxy group.

The MUSE observations used here were of short exposure time and taken in relatively poor conditions, but we have shown that a considerable amount of useful information may still be gleaned from it. Indeed based on these observations of a single galaxy group, it is clear that by repeating this work on multiple sightlines intersecting galaxy groups, we may statistically probe the group environment and disentangle the complex relationship between absorbers in the line of sight of distant quasars and the large scale structure as traced by the galaxy population in a highly efficient manner. Indeed, by comparing to existing dynamical$+$QSO absorption line studies of `isolated' galaxies \citep[e.g.][]{2011ApJ...733..105K,2015ApJ...812...83N}, differences between the halo gas of individual galaxies and that of galaxy groups may be investigated.

Taking the nearest galaxy in impact parameter ($b=49$~kpc) to the detected absorber system presents no obvious simple connection between the two. The velocity offset between the two would seem to be inconsistent with a co-rotating disk/halo model, especially given the low-luminosity (and likely low mass) of the galaxy. An outflow model associated with this nearest galaxy also appears unlikely given the low luminosity and low-SFR of the galaxy. The most plausible association in terms of an outflow model would be with the most highly star-forming galaxy detected in the galaxy group, however the galaxy impact parameter of $\approx101$~kpc would make this scenario seem extremely unlikely.

We find instead that the scenario most favoured by the data is that the absorber traces cool gas in the intra-group medium. The gas appears most likely to be multiple cool gas components potentially accreting into the group halo alongside individual group galaxies and currently existing at some stage in the process of mixing to form an intra-group medium, i.e. the cool gas components most likely originated in the original galaxy CGMs (via previous outflow or accretion) and are now entering the group environment alongside those parent galaxies.

Combining our analysis with previous detections of Mg~{\sc ii} absorbing gas likely associated with galaxy groups \citep[i.e.][]{2006MNRAS.368..341W,2010MNRAS.406..445K,2013MNRAS.432.1444G} provides a sample of four Mg~{\sc ii} absorber-group associations. From this sample we find a range of Mg~{\sc ii} equivalent widths of $W_r(2796)\sim0.7-4.2$~\AA. A range of other low-ionization absorption species are detected at the Mg~{\sc ii} absorber redshift in our own and other absorber-group associations including Ca~{\sc ii}, Fe~{\sc ii}, Mg~{\sc i}, Mn~{\sc ii}, Na~{\sc i}, and Ti~{\sc ii}.

In summary, we find that the association between galaxies and low-ionization absorbers can be an extremely complex one in which simply taking the nearest detected galaxy can be misleading, whilst we argue that discarding absorber-galaxy pairs found in group environments from analyses can exclude a large fraction of astrophysically interesting systems. Galaxy group systems are inherently complex, especially when considering the baryon content and structure, but contain a necessarily high fraction of the overall baryonic content of the Universe. Probing the gas in such systems via absorption line studies is thus important for understanding the broader topic of galaxy evolution and the role baryons play in this. 

\section*{Acknowledgements}

RMB, MF, and SLM acknowledge support from STFC (ST/L00075X/1). JPS gratefully acknowledges support from a Hintze Research Fellowship. NT acknowledges support from {\it CONICYT PAI/82140055}.

We thank M. T.~Murphy and S.~Kotus for making their combined UVES spectrum of HE0515-4414 available prior to publication. In addition we thank R. Bower and A. M. Swinbank for their contributions to this project and R. Crain, L. Parker, A. Meiksin and T. Ponman for their comments/discussion on this work. We also thank the anonymous referee for their comments and suggestions.

This research has made use of the NASA/IPAC Extragalactic Database (NED) which is operated by the Jet Propulsion Laboratory, California Institute of Technology, under contract with the National Aeronautics and Space Administration. In particular, we thank M.~Schmitz of IPAC who provided us with specific MRSS data not yet available through the public NED listings.

In the course of this work, we made use of Ankit Rohatgi's WebPlotDigitizer software\footnote{\url{http://arohatgi.info/WebPlotDigitizer/}}.




\bibliographystyle{mnras}
\bibliography{$HOME/Dropbox/lib/rmb} 




\appendix

\section{Other systems in the sightline}

Although our focus here is the low redshift galaxy group, we also note that the MUSE data provides redshifts for 1 lower and 6 higher redshift galaxies. We briefly note here the associations within $\Delta v\sim300$~km~s$^{-1}$ between these galaxies and sightline absorption systems.

The E230M STIS data covers a Ly$\alpha$ redshift range of $0.87<z<1.56$ and the detected galaxies that fall within the range are J051707-441036, J051705-441115 and J051710-441054.

J051707-441036 corresponds to two Ly$\alpha$ lines within $\lesssim300-400$~km~s$^{-1}$ with column densities of $N_{\rm HI}=10^{13.3}$~cm$^{-2}$ and $N_{\rm HI}=10^{15.1}$~cm$^{-2}$ \citep{2006A&A...458..427J}. Also a $N_{\rm CIV}\approx10^{14}$~cm$^{-2}$ C~{\sc iv} system lies within only a few km~s$^{-1}$ of the galaxy redshift \citep{2010ApJ...708..868C}.

J051705-441115 ($z=1.0183$) lies within $\Delta v\lesssim350$~km~s$^{-1}$ of 5 low column density ($N_{\rm HI}=10^{12-13}$~cm$^{-2}$) Ly$\alpha$ absorbers  \citep{2006A&A...458..427J}.

The original subject of the MUSE proposal was a sub-DLA at $z\approx1.15$ \citep[$N_{\rm HI}=10^{19.8}$~cm$^{-2}$,][]{2006A&A...458..427J} and this lies within $\approx200$~km~s$^{-1}$ of J051710-441054. We note that this galaxy is only weakly detected in our data however and requires further observations to ascertain a reliable redshift.

The remaining galaxies in our sample correspond to redshifts that have so far not been probed in Ly$\alpha$ for the background QSO. We therefore list any known metal lines at the galaxy redshifts.

J051710-441054 at $z=0.5181$ is found to lie within $\Delta v\approx100$~km~s$^{-1}$ of a $N_{\rm CIV}>10^{14.5}$~cm$^{-2}$ C~{\sc iv} system \citep{2010ApJ...708..868C}. J051710-441044 corresponds to a $N_{\rm CIV}\approx10^{13.5}$~cm$^{-2}$ and a $N_{\rm CIV}>10^{14.}$~cm$^{-2}$ at $\Delta v\approx60$~km~s$^{-1}$ and $\Delta v\approx200$~km~s$^{-1}$ respectively \citep{2010ApJ...708..868C}.

%


\bsp	
\label{lastpage}
\end{document}